# Binomial multichannel algorithm

E-mail: lanin99@mail.ru

*It is devoted to MVA anniversary (70+40)*

**Abstract**: The binomial multichannel algorithm is proposed. Some its properties are discussed.

## Introduction

The algorithm of universal compression (so-called MV2 algorithm) was firstly proposed in [1-2]. It is based on bit (binary) recording of file and represented an expressive example of multichannel crypto-algorithms. The theory of binomial radix with binary alphabet was stated in [3-4]. In this report the abovementioned ideas are used to construct the binomial multichannel algorithm and to discuss some its properties. This continuation of my works [5-6] in direction generalization of procedure of bit recording on other radix.

## Definitions

At the beginning I shall give some definitions.

**Bit** (short for **b**inary dig**it** - abbreviated **b**) - quantity which has equally only two values: a zero or a one.

**N-digit tuple** – a bit's sequence consisting of N bit.

**Binary binomial numbers** - the numerical function of a *n*-digit *k*-binomial radix with binary alphabet.

**N-digit alphabet ( $A_N$ )** - set of all various N-digit tuple.

**n-digit k-binomial alphabet ( $B_{k,n}$ )** - set of all various binary binomial numbers with parameters (n, k).

**Basic file** - a sequence of the N-digit alphabet ( $A_N$ ) without recurrence.

**Basic set** - a set of independent classes **X** and **Y** for given $N$.

**Length of a file** – amount of bit used for representation of a file.

**MV2 bit recording of N-digit alphabet ( $A_N$ )** - isomorphism between a set of the N-digit alphabet ( $A_N$ ) with constant length of a tuple **N** and finite set of alphabets $A_K$ with the **constant** length of a tuple $1 \leq K \leq N-1$.

**Binomial bit recording of N-digit alphabet ( $A_N$ )** - isomorphism between a set of the N-digit alphabet ( $A_N$ ) with constant length of a tuple **N** and finite set of alphabets $B_{k,n}$ with the **variable** length of a tuple $1 \leq K \leq N-1$.

**The coefficient of compression** $k_{\min} = \dfrac{L_2}{L_1}$ - the ratio of lengths for the basic file after $L_2$ and before $L_1$ bit recording of the N-digit alphabet ( $A_N$ )

## MV2-algorithm and its clones

My discussion will begin with the following equation:

$$2^N = \sum_{K=1}^{K=N-1} 2_N^K + 2 = 2^1 + \ldots + 2^{N-1} + 2. \quad (1)$$

It shows, that N-digit alphabet ($A_N$) can be divided into finite set of alphabets ($A_K$) with constant length of a tuple $1 \leq K \leq N-1$ plus two any elements from initial N-digit alphabet ($A_N$), i.e.

$$A_N = \sum_{K=1}^{K=N-1} A_K + a_N^1 + a_N^2, \quad (2)$$

where $a_K^i \in A_K$ and $i \in [1; 2^K]$.

According to this MV2 bit recording of N-digit alphabet ($A_N$) the initial file is divided on two parts and any of them has no neither statistical and nor semantic links with an initial file. The first part names a core (kernel) and arises from an initial file by bit recording (see equation (2)). The second part (so-called flag) gives us the information on the length of a tuple $K$ for each element. This information is the information on number of the used alphabet also. Because of obvious compression of an initial file in comparison with the core, all procedure can be repeated some times. These are a key moments of MV2 algorithm. For the best understanding its realization is displayed for the basic file of the alphabet $A_N$ on scheme N1.

| An initial (basic) file | Flag | | Core |
|---|---|---|---|
| 000..00000 | 1 | 000..0000_ | __.._____0 |
| 000..00001 | 1 | 000..0000_ | __.._____1 |
| 000..00010 | 1 | 000..000__ | __.._____00 |
| 000..00011 | 1 | 000..000__ | __.._____01 |
| 000..00100 | 1 | 000..000__ | __.._____10 |
| 000..00101 | 1 | 000..000__ | __.._____11 |
| 000..00110 | 1 | 000..00___ | __.._____000 |
| 000..00111 | 1 | 000..00___ | __.._____001 |
| 000..01000 | 1 | 000..00___ | __.._____010 |
| 000..01001 | 1 | 000..00___ | __.._____011 |
| 000..01010 | 1 | 000..00___ | __.._____100 |
| 000..01011 | 1 | 000..00___ | __.._____101 |
| 000..01100 | 1 | 000..00___ | __.._____110 |
| 000..01101 | 1 | 000..00___ | __.._____111 |
| ………….. | … | …………. | ………….. |
| 111111010 | 1 | 0__.._____ | _111..1100 |
| 111111011 | 1 | 0__.._____ | _111..1101 |
| 111111100 | 1 | 0__.._____ | _111..1110 |
| 111111101 | 1 | 0__.._____ | _111..1111 |
| 111111110 | 1 | ___.._____ | 0000..0000 |
| 111111111 | 1 | ___.._____ | 0000..0001 |

Scheme N1. Realization of MV2 algorithm for alphabets $A_N$ [1].

There is an opportunity to realize clones of MV2 algorithm - for it is necessary to use instead of the equation (2) the following equations:

$$A_N = \sum_{K=1}^{K=N} \sum_{i=1}^{i=M_K} a_K^i; \quad (3a)$$

$$\sum_{K=1}^{K=N} M_K = 2^N, \quad (3b)$$

where $M_K$ is quantity of elements of the alphabet ($A_K$).

It means, that some elements of the initial N-digit alphabet ($A_N$) with constant length $N$ (see equation (3a)) are replaced on the elements of alphabets ($A_K$) with a constant, but smaller length $1 \le K \le N-1$ at one restriction: the total of elements remains to constant (see equation (3b)). Each such clone is characterized by the coefficient of the compression $k_{min}^{clone}$ [5-6]:

$$k_{min}^{clone} = \frac{\sum_{K=1}^{K=N} K * M_K}{N * 2^N}.$$

It has the smallest value for MV2 algorithm:

$$k_{min}^{MV2} = \frac{2N + 2 \sum_{l=0}^{l=N-2}(l+1)*2^l}{N * 2^N}.$$

**Binomial radix**

The binomial radix is the radix with binomial weights, i.e. the numerical function of a *n*-digit *k*-binomial radix defining a quantitative equivalent of considered numbers $B_j = b_1 b_2 \ldots b_r$ by following expression

$$B_j = b_1 b_2 \ldots b_j \ldots b_r = \sum_{j=1}^{j=r} b_j C_{n-j}^{k-q_j}, \qquad (4)$$

where $q_j = \sum_{t=1}^{t=j-1} b_t$; n, k - parameters of a radix; $j = [1, C_n^k]$.

Further the binary alphabet will be used, i.e. $b_r \in B = \{0;1\}$. The binary binomial numbers are characterized by parameters (n, k); they have variable length $r$ ($1 \le r \le n-1$), contain or *k* units and thus come to an end on 1, or have ($n - k$) zero and thus come to an end on 0, i.e. there are following restrictions:

$$\begin{cases} k \le r \le n-1 \\ q = k \\ b_r = 1 \end{cases} \qquad (5a)$$

and

$$\begin{cases} n-k = r-q \\ 0 \le q \le k-1 \\ b_r = 0 \end{cases} \qquad (5b)$$

According to [3-4] above-stated restrictions (5a, 5b) divide binary binomial numbers $B_j \in B_{k,n}$ into two not crossed classes **X** and **Y**; $B_{k,n} = X \cup Y$; $X \cap Y = \emptyset$. The first class **X** contains *k* units and *l* zero, $0 \le l \le n-k-1$, and the second class **Y** – ($n-k$) zero and *q* units, $0 \le q \le k-1$. The range of the binomial radix represents binomial factor. As the concrete example the n-digit k-binomial alphabet ($B_{k,n}$) with some values of parameters (n, k) is shown in table N1.

This information also is all necessary for the further consideration.

## Binomial multichannel algorithm and its properties

Let's recollect that my discussion began with the equation (1):

$$2^N = \sum_{K=1}^{K=N-1} 2^K_N + 2 = 2^1 + \ldots + 2^{N-1} + 2.$$

However this equation can be transformed to other, combinatorial form:

$$2^N = (1+1)^N = \sum_{k=0}^{k=N} C_N^K = C_N^0 + C_N^1 + \ldots + C_N^{N-1} + C_N^N. \quad (6)$$

It shows that the N-digit alphabet ($A_n$) also can be divided into finite set of N-digit K-binomial alphabets ($B_{K,N}$), i.e.

$$A_N = \sum_{K=1}^{K=N} B_{K,N} \equiv \sum_{k=1}^{k=N-1}(X_k + Y_k) + B_{N,N} ; \quad (7a)$$

$$B_{N,N} \equiv \{0;1\}. \quad (7b)$$

Here entered the N-digit N-binomial alphabet ($B_{N,N}$) has only two values (see equation (7b)) and represents the single element from $C_N^0$ plus the single element $C_N^N$ according to equation (6). Similarly **MV2 bit recording of N-digit alphabet ($A_n$)** (see equation (2)) transformation (7a, 7b) will name as **binomial bit recording of N-digit alphabet ($A_n$)**. After binomial bit recording the initial file is divided by similar way on two parts and any of them has no neither statistical and nor semantic links with an initial file. However the second part (flag) gives us only the information concerning number of the used alphabet here, but not concerning the length of a tuple *K* for each element. Let's remind that the elements of N-digit K-binomial alphabets ($B_{K,N}$) have the variable length. Because of obvious compression of an initial file in comparison with the core, all procedure can be repeated some times.

From equality $C_n^k = C_n^{n-k}$ following communication between the classes **X** and **Y** in various alphabets $B_{k,n}$ can be received by operation P:

$$X_k = PY_{n-k} = \overline{Y_{n-k}}, \quad (8)$$

where $0 = P1 = \bar{1}, 1 = P0 = \bar{0}$.

This fact allows to change equation (7a) to next form:

$$A_N = \sum_{K=1}^{K=N} B_{K,N} \equiv \sum_{k=1}^{k=N-1}(X_k + Y_k) + B_{N,N} = \sum_{k=1}^{k=N-1}(X_k + \overline{X_k}) + B_{N,N} ; \quad (9)$$

and to introduce naturally such additional attribute as value of operation P for each element $X_k$. In this case the third additional, also not having semantic loading of the initial text a file - a flag N2 appears. The given realization is similar to the realization of second clone for MV2 algorithm in [6].

The coefficient of the compression $k_{min}^{bin}$ for binomial algorithm is equaled:

$$k_{min}^{bin} = \frac{2 + \sum_{l=0}^{l=N-2}(l+1)*C_N^{l+2}}{N*2^N}.$$

Also for the best understanding the realization of my algorithm is displayed for the basic file on schemes N2-7 and the values of the coefficient of the compression $k_{\min}^{bin}$ ($k_{\min}^{MV2}$) are given in table N2 only for $n \leq 8$. There is very big size of schemes for other values $n$ and because of the lack of a place they are not shown. Therefore only the basic set showed on the schemes 7;8 for a case $n \equiv 7;8$. For comparison with MV2 algorithm its core is given extreme by right column after a black line on all schemes.

There is no a direct opportunity to realize clones of my algorithm – the limited set of the elements of the binomial alphabets $B_{k,n}$ with variable length of a tuple exists. But there is the possibility to realize the mixed clones or a clones as with elements of the constant length of a tuple, belonging alphabets ($A_K$), and with elements of the variable length of a tuple, belonging binomial alphabets $B_{K,N}$.

**Conclusion**

We have proposed the binomial multichannel algorithm and have given obvious examples of its realization for $n \leq 8$. The generalization of considered algorithm on any values $n$ is simple. Two forms of flag are stated. The remark about symmetry $C_n^k = C_n^{n-k}$ or the operation P allows to add a new attribute as new channel. In difference from MV2 algorithm in my item the core is less and the flag shows only the number of a considered class. The coefficient of the compression $k_{\min}^{bin}$ for binomial algorithm and the mixed clones are discussed.

**Acknowledgments**

...
The author thanks Prof. Dr. Stephan Olariu and Doru Tiliute, "Stefan Cel Mare" University for kind hospitality at the IWISWN, Suceava, where the final version of the present work was written.

| N | K | Class $X_k \in X$ | Class $Y_k \in Y$ |
|---|---|---|---|
| 2 | 1 | $X_1 = \overline{Y_1} = \{1\}$ | $Y_1 = \{0\}$ |
| 3 | 1 | $X_1 = \overline{Y_2} = \{1; 01\}$ | $Y_1 = \{00\}$ |
| 3 | 2 | $X_2 = \overline{Y_1} = \{11\}$ | $Y_2 = \{0; 10\}$ |
| 4 | 1 | $X_1 = \overline{Y_3} = \{1; 01; 001\}$ | $Y_1 = \{000\}$ |
| 4 | 2 | $X_2 = \overline{Y_2} = \{11; 011; 101\}$ | $Y_2 = \{00; 100; 010\}$ |
| 4 | 3 | $X_3 = \overline{Y_1} = \{111\}$ | $Y_3 = \{0; 10; 110\}$ |
| 5 | 1 | $X_1 = \overline{Y_4} = \{1; 01; 001; 0001\}$ | $Y_1 = \{0000\}$ |
| 5 | 2 | $X_2 = \overline{Y_3} = \{11; 011; 101; 0011; 0101; 1001\}$ | $Y_2 = \{000; 1000; 0100; 0010\}$ |
| 5 | 3 | $X_3 = \overline{Y_2} = \{111; 0111; 1011; 1101\}$ | $Y_3 = \{00; 100; 010; 1100; 1010; 0110\}$ |
| 5 | 4 | $X_4 = \overline{Y_1} = \{1111\}$ | $Y_4 = \{0; 10; 110; 1110\}$ |
| 6 | 1 | $X_1 = \overline{Y_5} = \{1; 01; 001; 0001; 00001\}$ | $Y_1 = \{00000\}$ |
| 6 | 2 | $X_2 = \overline{Y_4} = \begin{Bmatrix} 11; 011; 101; 0011; 0101; 1001; \\ 00011; 00101; 01001; 10001 \end{Bmatrix}$ | $Y_2 = \{0000; 10000; 01000; 00100; 00010\}$ |
| 6 | 3 | $X_3 = \overline{Y_3} = \begin{Bmatrix} 111; 0111; 1011; 1101; 00111; \\ 01011; 01101; 10011; 10101; 11001 \end{Bmatrix}$ | $Y_3 = \begin{Bmatrix} 000; 1000; 0100; 0010; 11000; \\ 10100; 10010; 01100; 01010; 00110 \end{Bmatrix}$ |
| 6 | 4 | $X_4 = \overline{Y_2} = \{1111; 01111; 10111; 11011; 11101\}$ | $Y_4 = \begin{Bmatrix} 00; 100; 010; 11100; 1010; 0110; \\ 11100; 11010; 10110; 01110 \end{Bmatrix}$ |
| 6 | 5 | $X_5 = \overline{Y_1} = \{11111\}$ | $Y_5 = \{0; 10; 110; 1110; 11110\}$ |
| 7 | 1 | $X_1 = \overline{Y_6} = \{1; 01; 001; 0001; 00001; 000001\}$ | $Y_1 = \{000000\}$ |
| 7 | 2 | $X_2 = \overline{Y_5} = \begin{Bmatrix} 11; 011; 101; 0011; 0101; 1001; 100001; \\ 00011; 00101; 01001; 10001; \\ 000011; 000101; 001001; 010001; \end{Bmatrix}$ | $Y_2 = \begin{Bmatrix} 00000; 100000; 010000; \\ 001000; 000100; 000010 \end{Bmatrix}$ |
| 7 | 3 | $X_3 = \overline{Y_4} = \begin{Bmatrix} 111; 011; 1011; 1101; 00111; 01011; \\ 01101; 10011; 10101; 11001; 000111; \\ 001011; 010011; 100011; 001101; \\ 010101; 011001; 100101; 101001; 110001 \end{Bmatrix}$ | $Y_3 = \begin{Bmatrix} 0000; 10000; 01000; 00100; 00010; \\ 110000; 101000; 100100; 100010; 011000; \\ 010100; 001100; 010010; 001010; 000110 \end{Bmatrix}$ |
| 7 | 4 | $X_4 = \overline{Y_3} = \begin{Bmatrix} 1111; 01111; 10111; 11011; 11101; \\ 001111; 010111; 011011; 011101; 100111; \\ 101011; 110011; 101101; 110101; 111001 \end{Bmatrix}$ | $Y_4 = \begin{Bmatrix} 000; 1000; 0100; 0010; 11000; 10100; \\ 10010; 01100; 01010; 00110; 111000; \\ 110100; 101100; 011100; 110010; \\ 101010; 100110; 011010; 010110; 001110 \end{Bmatrix}$ |
| 7 | 5 | $X_5 = \overline{Y_2} = \begin{Bmatrix} 11111; 011111; 101111; \\ 110111; 111011; 111101 \end{Bmatrix}$ | $Y_5 = \begin{Bmatrix} 00; 100; 010; 1100; 1010; 0110; 011110; \\ 11100; 11010; 10110; 01110; \\ 111100; 111010; 110110; 101110 \end{Bmatrix}$ |
| 7 | 6 | $X_6 = \overline{Y_1} = \{111111\}$ | $Y_6 = \{0; 10; 110; 1110; 11110; 111110\}$ |

Table N1. The n-digit k-binomial alphabet ( $B_{k,n}$ ).

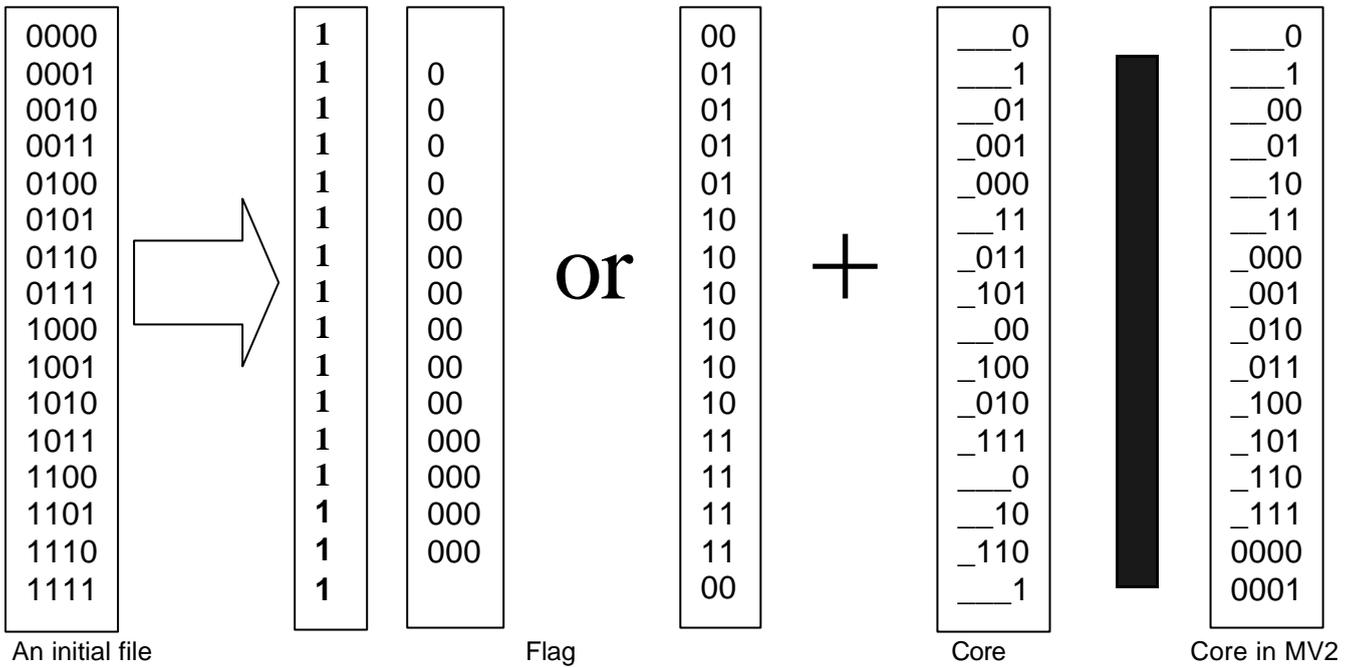

Scheme N2. Realization of binomial multichannel algorithm (n=4).

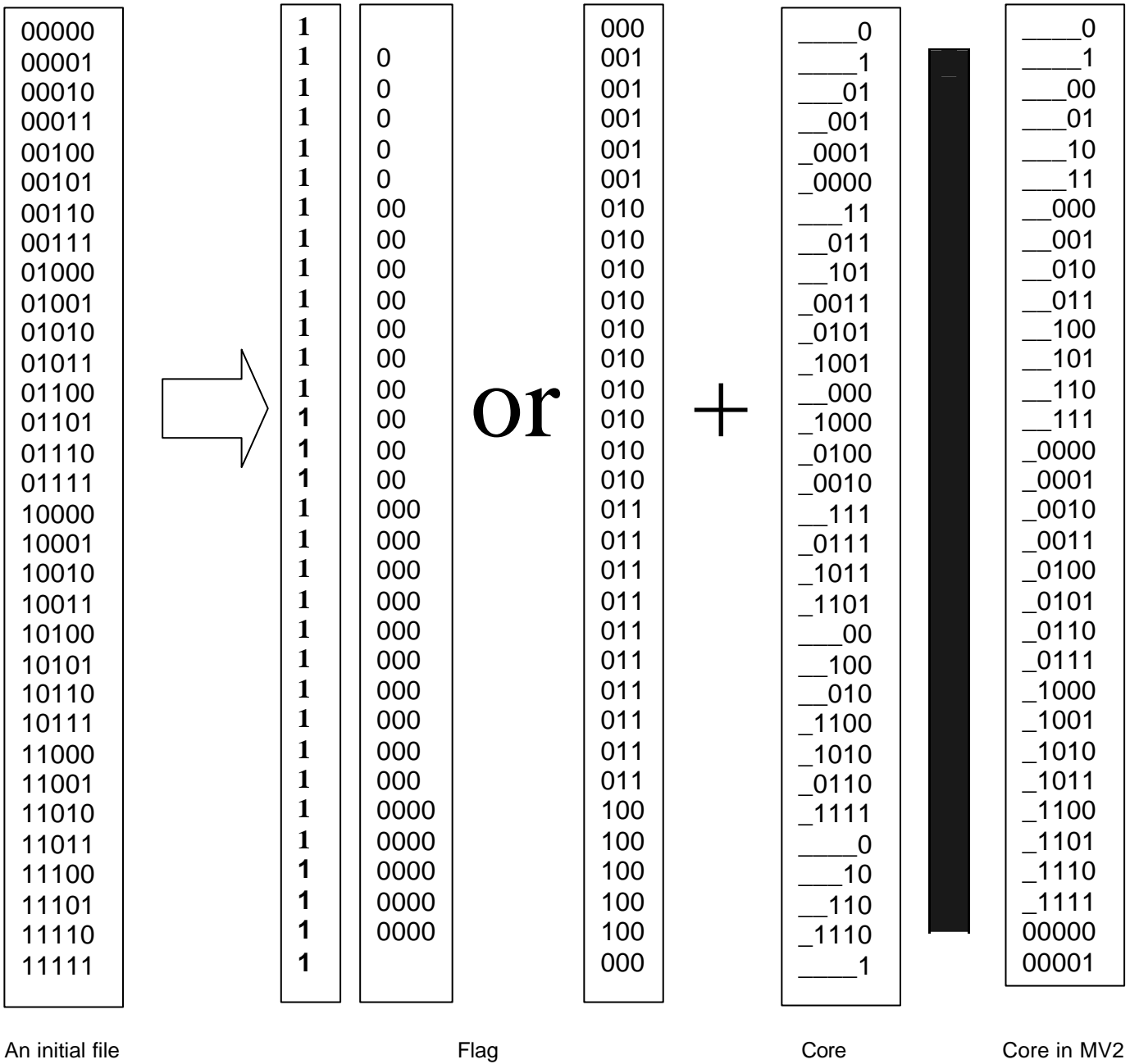

Scheme N3. Realization of binomial multichannel algorithm (n=5).

| An initial file | Flag | | Core | Core in MV2 |
|---|---|---|---|---|
| 000000 | 1 | | ____0 | ____0 |
| 000001 | 1 | 0 | ____1 | ____1 |
| 000010 | 1 | 0 | ____01 | ____00 |
| 000011 | 1 | 0 | ___001 | ____01 |
| 000100 | 1 | 0 | ___0001 | ____10 |
| 000101 | 1 | 0 | __00001 | ____11 |
| 000110 | 1 | 0 | __00000 | ____000 |
| 000111 | 1 | 00 | ____11 | ____001 |
| 001000 | 1 | 00 | ___011 | ____010 |
| 001001 | 1 | 00 | ___101 | ____011 |
| 001010 | 1 | 00 | __0011 | ____100 |
| 001011 | 1 | 00 | __0101 | ____101 |
| 001100 | 1 | 00 | __1001 | ____110 |
| 001101 | 1 | 00 | _00011 | ____111 |
| 001110 | 1 | 00 | _00101 | ___0000 |
| 001111 | 1 | 00 | _01001 | ___0001 |
| 010000 | 1 | 00 | _10001 | ___0010 |
| 010001 | 1 | 00 | __0000 | ___0011 |
| 010010 | 1 | 00 | _10000 | ___0100 |
| 010011 | 1 | 00 | _01000 | ___0101 |
| 010100 | 1 | 00 | _00100 | ___0110 |
| 010101 | 1 | 00 | _00010 | ___0111 |
| 010110 | 1 | 000 | ____111 | ___1000 |
| 010111 | 1 | 000 | __0111 | ___1001 |
| 011000 | 1 | 000 | __1011 | ___1010 |
| 011001 | 1 | 000 | __1101 | ___1011 |
| 011010 | 1 | 000 | _00111 | ___1100 |
| 011011 | 1 | 000 | _01011 | ___1101 |
| 011100 | 1 | 000 | _01101 | ___1110 |
| 011101 | 1 | 000 | _10011 | ___1111 |
| 011110 | 1 | 000 | _10101 | __00000 |
| 011111 | 1 | 000 | _11001 | __00001 |
| 100000 | 1 | 000 | ___000 | __00010 |
| 100001 | 1 | 000 | __1000 | __00011 |
| 100010 | 1 | 000 | __0100 | __00100 |
| 100011 | 1 | 000 | __0010 | __00101 |
| 100100 | 1 | 000 | _11000 | __00110 |
| 100101 | 1 | 000 | _10100 | __00111 |
| 100110 | 1 | 000 | _10010 | __01000 |
| 100111 | 1 | 000 | _01100 | __01001 |
| 101000 | 1 | 000 | _01010 | __01010 |
| 101001 | 1 | 000 | _00110 | __01011 |
| 101010 | 1 | 0000 | __1111 | __01100 |
| 101011 | 1 | 0000 | _01111 | __01101 |
| 101100 | 1 | 0000 | _10111 | __01110 |
| 101101 | 1 | 0000 | _11011 | __01111 |
| 101110 | 1 | 0000 | _11101 | __10000 |
| 101111 | 1 | 0000 | ____00 | __10001 |
| 110000 | 1 | 0000 | ___100 | __10010 |
| 110001 | 1 | 0000 | ___010 | __10011 |
| 110010 | 1 | 0000 | ___1100 | __10100 |
| 110011 | 1 | 0000 | ___1010 | __10101 |
| 110100 | 1 | 0000 | ___0110 | __10110 |
| 110101 | 1 | 0000 | _11100 | __10111 |
| 110110 | 1 | 0000 | _11010 | __11000 |
| 110111 | 1 | 0000 | _10110 | __11001 |
| 111000 | 1 | 0000 | _01110 | __11010 |
| 111001 | 1 | 00000 | _11111 | __11011 |
| 111010 | 1 | 00000 | ____0 | __11100 |
| 111011 | 1 | 00000 | ____10 | __11101 |
| 111100 | 1 | 00000 | ___110 | __11110 |
| 111101 | 1 | 00000 | __1110 | __11111 |
| 111110 | 1 | 00000 | _11110 | 000000 |
| 111111 | 1 | | ____1 | 000001 |

Scheme N4. Realization of binomial multichannel algorithm (n=6).

| An initial file | Flag | | Core | Core in MV2 |
|---|---|---|---|---|
| 000 | 1 | | __0 | __0 |
| 001 | 1 | 0 | __1 | __1 |
| 010 | 1 | 0 | _01 | _00 |
| 011 | 1 | 0 | _00 | _01 |
| 100 | 1 | 00 | _11 | _10 |
| 101 | 1 | 00 | __0 | _11 |
| 110 | 1 | 00 | _10 | 000 |
| 111 | 1 | | __1 | 001 |

Scheme N5. Realization of binomial multichannel algorithm (n=3).

| An initial file | | Flag | | or | Core | + | ■ | Core in MV2 |
|---|---|---|---|---|---|---|---|---|
| 00 01 10 11 | ⇒ | 1 1 1 1 | 0 0 | or | 00 01 01 00 | + | | _0 _1 00 01 |

Scheme N6. Realization of binomial multichannel algorithm (n=2).

| An initial file | | Flag | | or | Core | + | ■ | Core in MV2 |
|---|---|---|---|---|---|---|---|---|
| 0000000 0000001 0000010 0000011 0000100 0000101 0000110 0000111 0001000 0001001 0001010 0001011 0001100 0001101 0001110 0001111 0010000 0010001 0010010 0010011 0010100 0010101 0010110 0010111 0011000 0011001 0011010 0011011 0011100 0011101 0011110 0011111 0100000 0100001 0100010 0100011 0100100 0100101 0100110 0100111 0101000 0101001 0101010 0101011 0101100 0101101 0101110 0101111 0110000 0110001 0110010 0110011 0110100 0110101 0110110 0110111 0111000 0111001 0111010 0111011 0111100 0111101 0111110 0111111 ............ 1111111 | ⇒ | 1 1 1 1 1 1 1 1 1 1 1 1 1 1 1 1 1 1 1 1 1 1 1 1 1 1 1 1 1 1 1 1 1 1 1 1 1 1 1 1 1 1 1 1 1 1 1 1 1 1 1 1 1 1 1 1 1 1 1 1 1 1 1 1 ... 1 | 0 0 0 0 0 0 0 0 00 00 00 00 00 00 00 00 00 00 00 00 00 00 00 00 00 00 00 00 00 000 000 000 000 000 000 000 000 000 000 000 000 000 000 000 000 000 000 000 000 000 000 000 000 000 000 000 000 000 000 000 000 000 000 000 .... | or | 000 001 001 001 001 001 001 001 010 010 010 010 010 010 010 010 010 010 010 010 010 010 010 010 010 010 010 010 010 011 011 011 011 011 011 011 011 011 011 011 011 011 011 011 011 011 011 011 011 011 011 011 011 011 011 011 011 011 011 011 011 011 011 011 .... 000 | + | ______0 ______1 _____01 ____001 ___0001 __00001 _000001 _000000 ____11 ____011 ____101 ___0011 ___0101 ___1001 __00011 __00101 __01001 __10001 _000011 _000101 _001001 _010001 _100001 __00000 _100000 _010000 _001000 _000100 _000010 ____111 ___0111 ___1011 ___1101 __00111 __01011 __01101 __10011 __10101 __11001 _000111 _001011 _010011 _100011 _001101 _010101 _011001 _100101 _101001 _110001 ___0000 __10000 __01000 __00100 __00010 _110000 _101000 _100100 _100010 _011000 _010100 _001100 _010010 _001010 _000110 ............ ______1 | | ______0 ______1 _____00 _____01 _____10 _____11 ____000 ____001 ____010 ____011 ____100 ____101 ____110 ____111 ___0000 ___0001 ___0010 ___0011 ___0100 ___0101 ___0110 ___0111 ___1000 ___1001 ___1010 ___1011 ___1100 ___1101 ___1110 ___1111 __00000 __00001 __00010 __00011 __00100 __00101 __00110 __00111 __01000 __01001 __01010 __01011 __01100 __01101 __01110 __01111 __10000 __10001 __10010 __10011 __10100 __10101 __10110 __10111 __11000 __11001 __11010 __11011 __11100 __11101 __11110 __11111 _000000 _000001 ............ 1111111 |

Scheme N7 Realization of binomial multichannel algorithm (n=7).

| An initial file | | Flag | | Core | | Core in MV2 |
|---|---|---|---|---|---|---|
| 00000000 | | | 000 | _______0 | | _______0 |
| 00000001 | 1 | 0 | 001 | _______1 | | _______1 |
| 00000010 | 1 | 0 | 001 | ______01 | | ______00 |
| 00000011 | 1 | 0 | 001 | _____001 | | ______01 |
| 00000100 | 1 | 0 | 001 | ____0001 | | ______10 |
| 00000101 | 1 | 0 | 001 | ___00001 | | ______11 |
| 00000110 | 1 | 0 | 001 | __000001 | | _____000 |
| 00000111 | 1 | 0 | 001 | _0000001 | | _____001 |
| 00001000 | 1 | 00 | 010 | ______11 | | _____010 |
| 00001001 | 1 | 00 | 010 | _____011 | | _____011 |
| 00001010 | 1 | 00 | 010 | _____101 | | _____100 |
| 00001011 | 1 | 00 | 010 | ____0011 | | _____101 |
| 00001100 | 1 | 00 | 010 | ____0101 | | _____110 |
| 00001101 | 1 | 00 | 010 | ____1001 | | _____111 |
| 00001110 | 1 | 00 | 010 | ___00011 | | ____0000 |
| 00001111 | 1 | 00 | 010 | ___00101 | | ____0001 |
| 00010000 | 1 | 00 | 010 | ___01001 | | ____0010 |
| 00010001 | 1 | 00 | 010 | ___10001 | | ____0011 |
| 00010010 | 1 | 00 | 010 | __000011 | | ____0100 |
| 00010011 | 1 | 00 | 010 | __000101 | | ____0101 |
| 00010100 | 1 | 00 | 010 | __001001 | | ____0110 |
| 00010101 | 1 | 00 | 010 | __010001 | | ____0111 |
| 00010110 | 1 | 00 | 010 | __100001 | | ____1000 |
| 00010111 | 1 | 00 | 010 | _0000011 | | ____1001 |
| 00011000 | 1 | 00 | 010 | _0000101 | | ____1010 |
| 00011001 | 1 | 00 | 010 | _0001001 | | ____1011 |
| 00011010 | 1 | 00 | 010 | _0010001 | | ____1100 |
| 00011011 | 1 | 00 | 010 | _0100001 | | ____1101 |
| 00011100 | 1 | 00 | 010 | | _1000101 | ____1110 |
| 00011101 | 1 | 000 | 011 | _____111 | _1001001 | ____1111 |
| 00011110 | 1 | 000 | 011 | ____0111 | _1010001 | ___00000 |
| 00011111 | 1 | 000 | 011 | ____1011 | _1100001 | ___00001 |
| 00100000 | 1 | 000 | 011 | ____1101 | _0001111 | ___00010 |
| 00100001 | 1 | 000 | 011 | ___00111 | _0010111 | ___00011 |
| 00100010 | 1 | 000 | 011 | ___01011 | _0100111 | ___00100 |
| 00100011 | 1 | 000 | 011 | ___01101 | _1000111 | ___00101 |
| 00100100 | 1 | 000 | 011 | ___10011 | _0011011 | ___00110 |
| 00100101 | 1 | 000 | 011 | ___10101 | _0011101 | ___00111 |
| 00100110 | 1 | 000 | 011 | ___11001 | _0101011 | ___01000 |
| 00100111 | 1 | 000 | 011 | __000111 | _0101101 | ___01001 |
| 00101000 | 1 | 000 | 011 | __001011 | _0110011 | ___01010 |
| 00101001 | 1 | 000 | 011 | __010011 | _0110101 | ___01011 |
| 00101010 | 1 | 000 | 011 | __100011 | _0111001 | ___01100 |
| 00101011 | 1 | 000 | 011 | __001101 | _1001011 | ___01101 |
| 00101100 | 1 | 000 | 011 | __010101 | _1001101 | ___01110 |
| 00101101 | 1 | 000 | 011 | __011001 | _1010011 | ___01111 |
| 00101110 | 1 | 000 | 011 | __100101 | _1010101 | ___10000 |
| 00101111 | 1 | 000 | 011 | __101001 | _1011001 | ___10001 |
| 00110000 | 1 | 000 | 011 | __110001 | _1100011 | ___10010 |
| 00110001 | 1 | 000 | 011 | _0000111 | _1100101 | ___10011 |
| 00110010 | 1 | 000 | 011 | _0001011 | _1101001 | ___10100 |
| 00110011 | 1 | 000 | 011 | _0010011 | _1110001 | ___10101 |
| 00110100 | 1 | 000 | 011 | _0100011 | _0011111 | ___10110 |
| 00110101 | 1 | 000 | 011 | _1000011 | _0101111 | ___10111 |
| 00110110 | 1 | 000 | 011 | _0001101 | _0110111 | ___11000 |
| 00110111 | 1 | 000 | 011 | _0010101 | _0111011 | ___11001 |
| 00111000 | 1 | 000 | 011 | _0011001 | _0111101 | ___11010 |
| 00111001 | 1 | 000 | 011 | _0100101 | _1001111 | ___11011 |
| 00111010 | 1 | 000 | 011 | _0101001 | _1010111 | ___11100 |
| 00111011 | 1 | 000 | 011 | _0110001 | _1011011 | ___11101 |
| 00111100 | 1 | 000 | 011 | | _1011101 | ___11110 |
| 00111101 | 1 | 000 | 011 | | _1100111 | ___11111 |
| 00111110 | 1 | 000 | 011 | | _1101011 | __000000 |
| 00111111 | 1 | 000 | 011 | | _1101101 | __000001 |
| 01000000 | 1 | 0000 | 100 | | _1110011 | __000010 |
| 01000001 | 1 | 0000 | 100 | | _1110101 | __000011 |
| 01000010 | 1 | 0000 | 100 | | _1111001 | __000100 |
| 01000011 | | 0000 | 100 | | _0111111 | ………… |
| ………… | ... | …. | …. | | _1011111 | 11111111 |
| 11111111 | 1 | | 000 | | _1101111 | |
| | | | | | _1110111 | |
| | | | | | _1111011 | |
| | | | | | _1111101 | |
| | | | | | _1111110 | |
| | | | | | ………… | |
| | | | | | _______1 | |

Scheme N8 Realization of binomial multichannel algorithm (n=8).

| N | $k_{min}^{bin}$ | $k_{min}^{MV2}$ |
|---|---|---|
| 2 | 0.5 | 0.75 |
| 3 | 0.5 | 0.6(6) |
| 4 | 0.5625 | 0.65625 |
| 5 | 0.625 | 0.675 |
| 6 | 0.677083(3) | 0.703125 |
| 7 | 0.71875 | 0,7321428571428571428571428 |
| 8 | 0,751953125 | 0,7587890625 |

Table N2. The coefficient of the compression $k_{min}^{bin}$ ($k_{min}^{MV2}$) for binomial (MV2) algorithm.